\documentclass[11pt]{article}
\usepackage{subfigure}
\usepackage{cancel}
\usepackage{cite}
\usepackage{lipsum}
\usepackage{amsmath}
\usepackage{indentfirst}
\usepackage{multirow}
\usepackage{amssymb}
\usepackage{feynmp}
\usepackage[utf8]{inputenc}
\usepackage{hyperref}

\usepackage{amsmath,amssymb,amsthm,amsxtra,overpic,bbm,bm,epsfig,ulem}
\usepackage{booktabs}
\usepackage{mathrsfs}
\usepackage{graphicx}
\usepackage{dcolumn}
\usepackage{bm}
\usepackage{amsmath}
\usepackage{slashed}  
\usepackage{setspace}
\usepackage{color}

\begin{document}
\unitlength=1mm
\begin{center}
	{\Large \bf Search for vector-like bottom quark via $Zb$ production at the LHC}
\end{center}

\vspace{0.5cm}

\begin{center}
{\bf Xue Gong \footnote{E-mail: gongxue422@163.com},  Chong-Xing Yue \footnote{E-mail: cxyue@lnnu.edu.cn}, Yu-Chen Guo \footnote{E-mail: ycguo@lnnu.edu.cn}} \\
{Department of Physics, Liaoning Normal University, Dalian 116029, China}
\end{center}

\vspace{4cm}
\begin{abstract}
Vector-like quarks (VLQs) are predicted in many new physics scenarios beyond the standard model (SM). We consider a search  strategy for vector-like bottom quark (VLQ-$B$) at the LHC in  a model independent fashion. Our analysis is based on single production of VLQ-$B$ with decaying to the $ Zb$ mode. The production cross sections and the signal significance $SS$ are calculated, and  the observability  of the VLQ-$B$ signal is investigated. Our numerical results show that the possible  signals of the VLQ-$B$ with electric charge of $Q_B= -1/3$, which is  the $SU(2)$ singlet,  might be detected via the process $pp \to B \to Z~(\to l^+ l^-) b$  at the upgraded LHC.

\vspace{10cm}	
\end{abstract}

\vspace{18cm}

\section*{\uppercase\expandafter{\romannumeral1}. Introduction}

The standard model (SM) can explain with great accuracy most of the available experimental phenomena including the newest data from the large hadron collider (LHC).
However, there are still some unexplained discrepancies and theoretical issues that the SM can not solve.
As we know, the SM  is a very successful gauge theory, which can naturally be extended through adding the form of  gauge symmetries with new matter fields.
The measurements of the Higgs boson mediated cross-sections\cite{Aad:2012tfa}  indicate that the existence of the fourth generation of chiral quarks is in disfavour \cite{Eberhardt:2012gv}. However, vector-like quarks (VLQs) are a class of interesting new particles, which mix with the SM quarks and can lead to rich phenomenology.

Unlike chiral quarks of the SM,  which get the mass terms through electroweak symmetry breaking, VLQs obtain the direct gauge invariant mass term of  the form $m\bar{\psi}\psi$.
This means that VLQs are not subject to the constraints from Higgs production and not excluded by precision measurements \cite{Eberhardt:2012sb}.
~Furthermore, if Higgs boson is a pseudo-Goldstone boson of a  spontaneously broken approximate symmetry group, the VLQ of around TeV mass will be presented in many models beyond the SM, which can provide a viable option to explain the observed lightness of the Higgs\cite{Perelstein:2003wd}.

~VLQs are proposed in several theoretical frameworks, like little Higgs\cite{ArkaniHamed:2002qy} and composite Higgs\cite{Dobrescu:1997nm}.
~In these new physics models, VLQs are colored spin-1/2 fermions, whose left- and right-handed components transform in the same way under the SM electroweak symmetry group\cite{delAguila:1982fs}.
VLQs appear as partners of the third generation of quarks. The electric charges of VLQs could be $Q_T= +2/3$, $Q_B= -1/3$, $Q_X= +5/3$, and $Q_Y= -4/3$. They might appear in $SU(2)$ singlets or multiplets.
~These new fermions have a common feature that they are assumed to decay to a SM quark with a SM gauge boson, or a Higgs boson.
Many phenomenological  studies for VLQs at the existing or future colliders have been made in literatures, for example see \cite{Fuks:2016ftf}.

Among various new particles, VLQs play an important role in terms of experimental effort.~By now, the experimental data corresponding to an integrated luminosity of
35-36 fb$^{-1}$ have been used to search for the various VLQ decay modes\cite{Sirunyan:2018omb,Aaboud:2018uek,Aaboud:2018xuw,Sirunyan:2017ynj,Sirunyan:2018fjh,Sirunyan:2018ncp,
ATLAS:2016ovj}. The most stringent bounds on the down-type VLQ (VLQ-$B$) decaying into $Zb$, $Wt$, $hb$ are 1.24 TeV from CMS \cite{Sirunyan:2018omb}  and 1.35 TeV from
ATLAS \cite{Aaboud:2018uek} experiments. Similarly,  considering the decay modes $ T \to Zt\;, W^{+}b\;, ht $ of up-type VLQ (VLQ-$T$), the most stringent bounds on the mass is 1.3 TeV and 1.43 TeV given by CMS \cite{Sirunyan:2018omb} and ATLAS \cite{Aaboud:2018xuw}, respectively.
~Since pair production of VLQs is induced by QCD interaction, these mass bounds are independent of the value of the mixing angles with the SM quarks.
~Single production of VLQs, whose rate is proportional to the mixing angles, has also been searched by the ATLAS and CMS experiments at the 13 TeV
LHC \cite{Sirunyan:2017ynj,Sirunyan:2018fjh,Sirunyan:2018ncp,ATLAS:2016ovj}. In the case of the single production of VLQ-$B$ decaying into $hb$, the upper limits on the product of
the cross section and the branching fraction vary from 1.28 to 0.07 pb for the VLQ-$B$ mass in the range 700 - 1800 GeV\cite{Sirunyan:2018fjh}.
 The CMS collaboration has also published a search for VLQ-$T$ in the final state $Zt$\cite{Sirunyan:2017ynj}, while a search in the final state $Wb$ is given in the 2015 ATLAS dataset\cite{ATLAS:2016ovj}. We expect that more information about VLQs can be obtained in future high energy collider experiments.

VLQs can be produced singly or in pairs at the LHC. The pair production process induced by gluon fusion is model-independent. For heavy VLQs, the single production process becomes more important because of weaker phase-space suppression\cite{DeSimone:2012fs}.  Although the exact details of the corresponding production mechanism are model-dependent, one can assume that VLQs mix with the SM quarks and general consider their single productions.  In this letter, we will take that the VLQ-$B$ with electric charge of $Q_B= -1/3$ is  the $SU(2)$ singlet, consider its single  productions at the 14 TeV LHC in a model independent way, and careful simulate the signals and SM backgrounds for the subprocess $bg\to B\to Zb$.

If working in the 5 Flavor Scheme, which assumes a non-zero $b$-quark parton distribution function (PDF)\cite{Figueroa:2018chn},  production of a $Z$ boson associated with a $b$ mainly proceeds via the subprocess $bg \to Zb$ in the SM. Refs.\cite{Campbell:2003dd,Figueroa:2018chn} have calculated the leading order of QCD corrections and  next-to-leading order (NLO) QCD corrections to its production cross section. In this work, we will first consider all possible single production channels of VLQ-$B$ at the  LHC in the case of assuming that the VLQ-$B$ mass is larger than the electroweak (EW) scale. Then we will focus our attention on the observability of the VLQ-$B$ signals by considering its contributions to the subprocess $bg\to Zb$  at the 14 TeV LHC.

The letter is organized as follows. On the basis of the most general Lagrangian, which is related to the couplings of VLQ-$B$ with the SM particles, in Section II we consider its various decays and single  productions at the LHC. The corresponding branching ratios and production cross sections are calculated. In Section III, we  give a detailed analysis of the relevant signals and backgrounds for the process $pp \rightarrow B\to Zb$. Our results are summarized in Section IV.

\section*{\uppercase\expandafter{\romannumeral2}. Decays and single productions of VLQ-$B$  at the LHC}

The couplings of VLQs to the SM  particles are uniquely fixed by gauge invariance \cite{Cabibbo:1983bk}. The most general Lagrangian describing the effective interaction of VLQ-$B$, can be given by \cite{20}
\begin{eqnarray}
\mathcal{L}&=&\frac{g_s}{2\Lambda}G_{\mu\nu}\overline{b}\sigma^{\mu\nu}\Big(\kappa^b_L P_L+\kappa^b_R P_R\Big)B+\frac{g_2}{\sqrt{2}}W_{\mu}^{-} \overline{t} \gamma^{\mu}\Big(Y_LP_L+Y_RP_R\Big)B \nonumber\\
&+&\frac{g_2}{2c_W}Z_{\mu}\overline{b}\gamma^{\mu}\Big(F_L P_L+F_R P_R\Big)B+\frac{m_b}{\upsilon}h\overline{b}\Big(y_L P_L+y_R P_R\Big)B+h.c.. \label{Lag1}
\end{eqnarray}
In above equation, we have abbreviated  $\cos\theta_W$ as $c_W$, and $\theta_W$ is the Weinberg angle. $\Lambda$ is the cutoff scale which is set to the VLQ-$B$  mass $M_B$. $g_s$ and $g_2$ are the strong coupling constant and the $SU(2)_L$ coupling constant, respectively. $G_{\mu\nu}$ is the field strength tensor of the gluon. $m_b$ is the bottom quark mass. $P_{L,R}$ are the normal chiral projection operators with $P_{L,R} = (1\mp \gamma_5)/2$. The factors $\kappa^b_{L,R}$, $Y_{L,R}$, $F_{L,R}$ and $y_{L,R}$ parameterize the chirality of the VLQ-$B$ couplings with the SM particles. For the singlet  VLQ-$B$ with electric charge of $Q_B= -1/3$, it only mixes with the left-handed bottom quark and there are  $Y_R \simeq 0, F_R \simeq 0$ and $y_R \simeq \frac{M_B}{m_b} y_L$ . In this letter, we only consider this kind of VLQ-$B$ and assume $Y_L =F_L = y_{L}=s_L$ with $s_L$ being the mixing angle, $\kappa^b_{L}= \kappa^b_{R}= \kappa^b$. The current high energy experimental data can give constraints on these free parameters. In our following numerical estimation, we will take a conservative range for the parameters $s_L= {\upsilon}/{M_B}$, where $\upsilon$ =246 GeV is the EW scale, and $0\leq \kappa^b\leq 0.5$ according to Refs.~\cite{20,ycx}.
\begin{figure}[htbp]
	\centering
	\subfigure{
		\begin{minipage}[b]{0.5\textwidth}
			\includegraphics[width=1\textwidth]{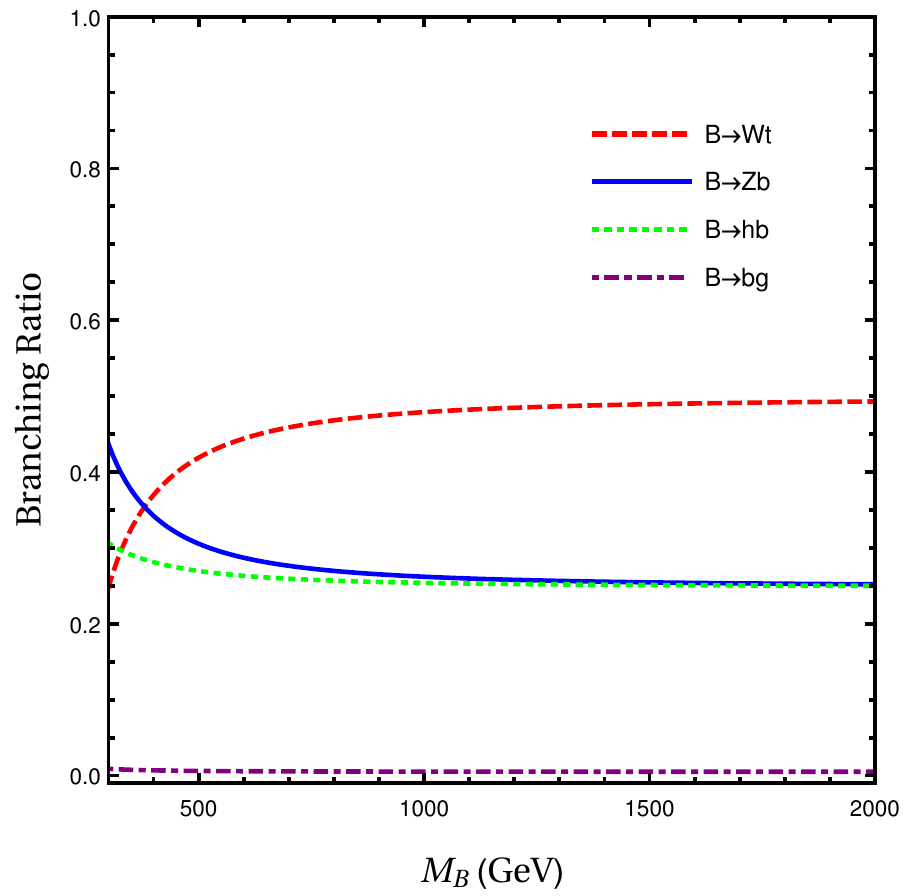}
		\end{minipage}
	} 	
	\caption{The branching ratios for the decay modes $Wt$, $Zb$, $hb$ and $bg$ as functions of the \hspace*{1.65cm} VLQ-$B$ mass $M_B$ for  $\kappa^b= 0.1$.}
	\label{fig1}
\end{figure}

From Eq. (\ref{Lag1}), we can see that the possible decay modes of VLQ-$B$ are $ ~Wt, ~Zb, ~hb$, and $bg$. The branching ratios for these decay channels  are plotted as functions of the mass parameter $M_B$ in Fig.~\ref{fig1} in the case of $\kappa^b= 0.1$. One can see from Fig.~\ref{fig1} that, for the relatively light  VLQ-$B$, the process $B \rightarrow Zb$ is the dominant decay channel, while for $M_B>1000GeV$ the branching ratio of the decay mode $Wt$ will increase to about 50\%, and there is $Br(B \rightarrow Zb) \simeq Br(B \rightarrow hb) $. Certainly, if we enhance the value of the parameter $\kappa^b$, the value of the branching ratio $Br(B \rightarrow bg)$ will be increased. It should be noted that Ref.~\cite{20} has studied the possibility of detecting VLQ-$B$ via the  decay channel $B \rightarrow Wt$ at the LHC.
Hence, in this letter, we only use the decay process $B \rightarrow Zb$ to study the possible signals of VLQ-$B$ at the LHC.
\begin{figure}[htbp]
	\centering
	\subfigure{
		\begin{minipage}[b]{0.55\textwidth}
			\includegraphics[width=1\textwidth]{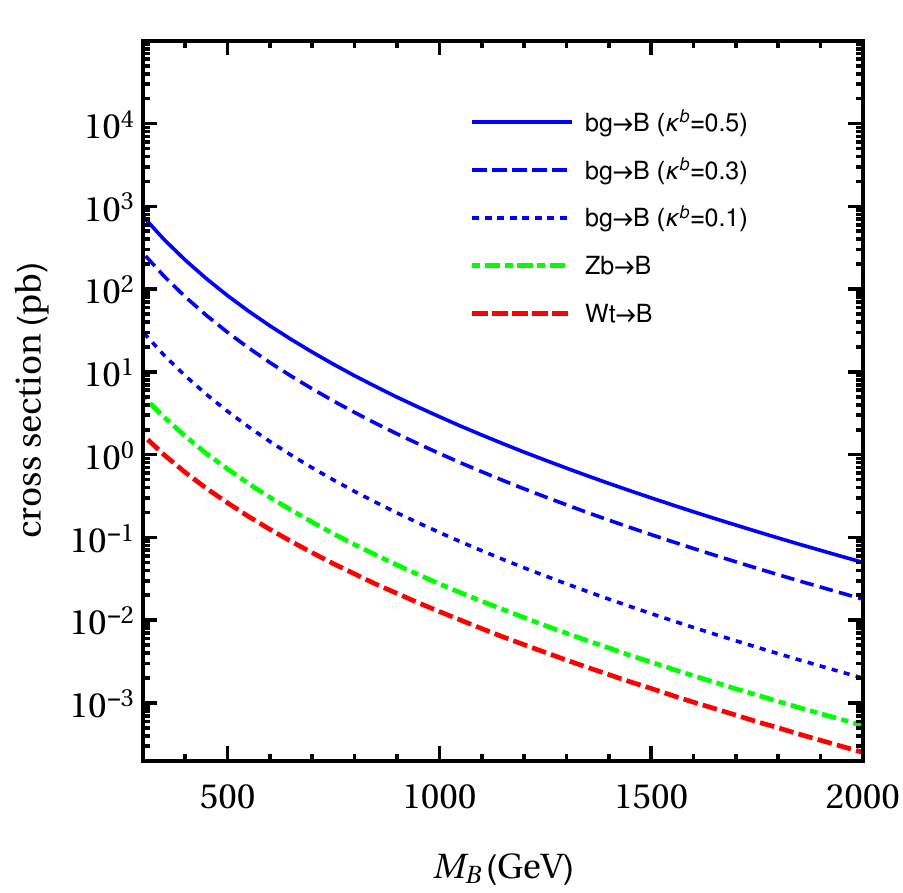}
		\end{minipage}
	}
	\caption{ The single production cross sections of VLQ-$B$ as functions of  $M_B$ with different  values \hspace*{1.65cm}of $\kappa^b$ at the 14TeV LHC.}
	\label{fig2}
\end{figure}

From above discussions we can see that VLQ-$B$ can be single  produced via $bg$, $Zb$ and $Wt$ fusions at the LHC.  It is obvious that the production cross section coming from $bg$ fusion depends on the free parameters $M_B$ and $\kappa^b$, while other production cross sections  only depend on the free parameter $M_B$ in the case of $s_L= {\upsilon}/{M_B}$. The relevant production cross sections are plotted in Fig.~\ref{fig2} as functions of  $M_B$  with different values of $\kappa^b$ at the 14TeV LHC. From Fig.~\ref{fig2} we can see that the production cross section of the subprocess  $bg \rightarrow B$ is larger than other two production channels  for $\kappa^b= 0.3 $ and $0.5$. For $\kappa^b= 0.1$, the production cross section induced by $Zb$ fusion  can compare with that from $bg$ fusion. However, comparing with the $Zb \rightarrow B$ production process, the $bg$ production channel less one $q$ jet in the final state and more easier to be analyzed. So we  study the possibility of detecting the signals of VLQ-$B$ via the subprocess $bg\to B\to Zb$ at the 14 TeV LHC.

\section*{III. Signal analysis and discovery potentiality}

After discussing all possible decays and single productions of VLQ-$B$ in Sec.II, in this section we will investigate  the signatures of VLQ-$B$ by considering its contributions to $Zb$ production at the 14 TeV LHC.  Ref.~\cite{ycx} has shown that the correction of VLQ-$B$ to $Zb$ production is comparable to  NLO QCD correction in wide range of the parameter space  and might be larger than the NLO QCD correction for some  special values of the free parameters. We will consider prospects for observing VLQ-$B$ at 14 TeV LHC in the process $pp\rightarrow B \rightarrow Zb$ with $Z$ decaying into two leptons. During our simulation procedure, we first implement the model with VLQ-$B$ into the FeynRules package~\cite{fr} to generate the model files in UFO~\cite{UFO} format. The calculation of cross sections and generation of parton level events are performed using  MadGraph5-aMC@NLO\cite{mg} with built-in parton distribution function NNPDF23~\cite{nnpdf}. Subsequently, the events pass through Pythia~\cite{pythia} and Delphes~\cite{delphes} which are employed for parton showering, hadronization and fast simulation of ATLAS detector. Finally, MadAnalysis5~\cite{md} is applied for data analysis and plotting.
\begin{figure}[htbp]
	\centering
	\subfigure{
		\begin{minipage}[b]{0.47\textwidth}
			\includegraphics[width=1\textwidth]{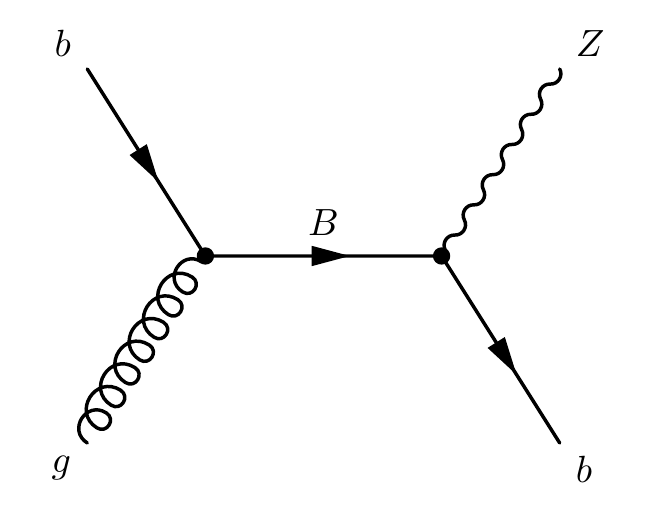}
		\end{minipage}
	}
	\caption{The Feynman diagram for the resonance production of VLQ-$B$  subsequently   decaying \hspace*{1.65cm}to $Zb$ at the LHC.}
	\label{fig3}
\end{figure}

\begin{figure}[htbp]
	\centering
	\subfigure{
		\begin{minipage}[b]{0.5\textwidth}
			\includegraphics[width=1\textwidth]{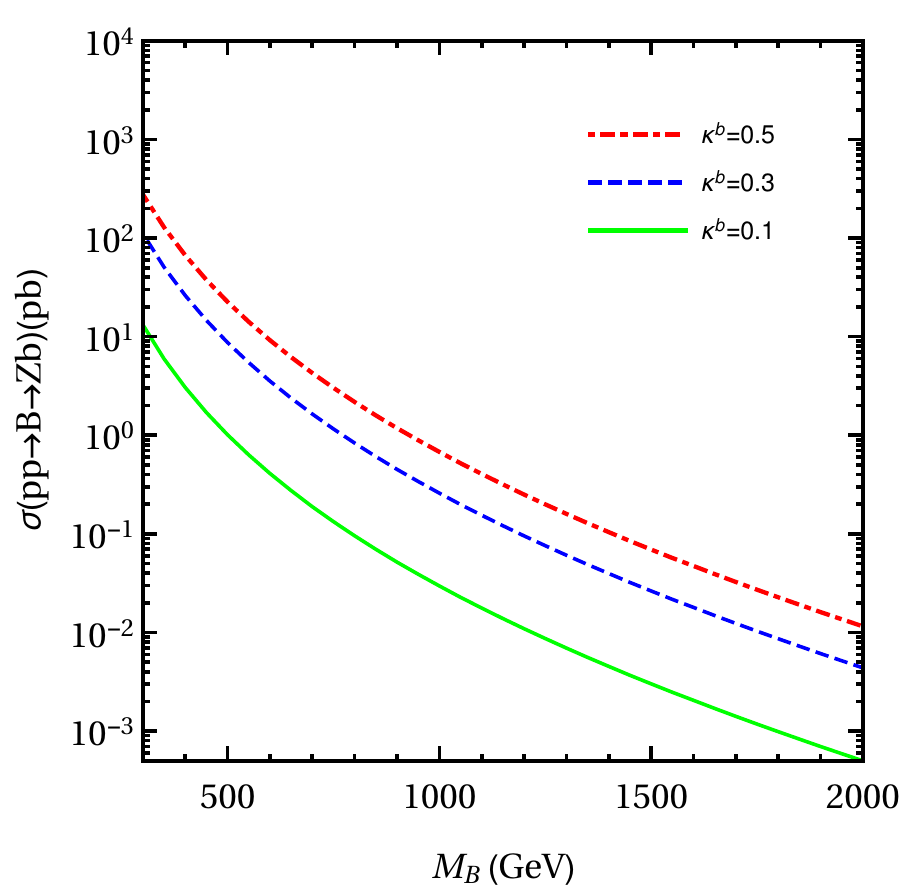}
		\end{minipage}
	}
	\caption{ The production cross section of the process $pp\rightarrow B \rightarrow Zb$ as a function of  $M_B$ for \hspace*{1.65cm} different values of  $\kappa^b$  at the 14 TeV LHC.}
	\label{fig4}
\end{figure}

VLQ-$B$ can be single produced via $bg$ fusion at the LHC and its subsequent decaying to $Zb$ can contribute to the process $pp\rightarrow Zb$. The relevant Feynman diagram is shown in Fig.~\ref{fig3}. The production cross section $ \sigma(pp\rightarrow B \rightarrow Zb )$ is plotted in Fig.~\ref{fig4} as a function of the  mass $M_B$ for different values of the parameter $\kappa^b$  at the 14 TeV LHC. For $0\leq\kappa^b \leq0.5$ and 300 GeV $<M_B<$ 2000 GeV, the values of the cross section  $ \sigma(pp\rightarrow B \rightarrow Zb )$ are in the range of $3.09\times10^{-4} \sim 276.72~  $pb. For $\kappa^b=0.1$ and $M_B$ = 800, 1000, 1200 GeV, their values are 0.10, 0.03, 0.01 $\rm pb$, respectively. Certainly, the value of the production cross section also depends on the PDFs of bottom quark and gluon. The cross section benefits from high gluon PDF but it is also suppressed by the PDF of bottom quark. Since in the single production the VLQ-$B$ mass might reach the full center-of-mass energy, single production can be more sensitive to heavy VLQ-$B$ than pair production.

\begin{figure}[htbp]
	\centering
	\subfigure[]{
		\begin{minipage}[b]{0.48\textwidth}
			\includegraphics[width=1.45\textwidth]{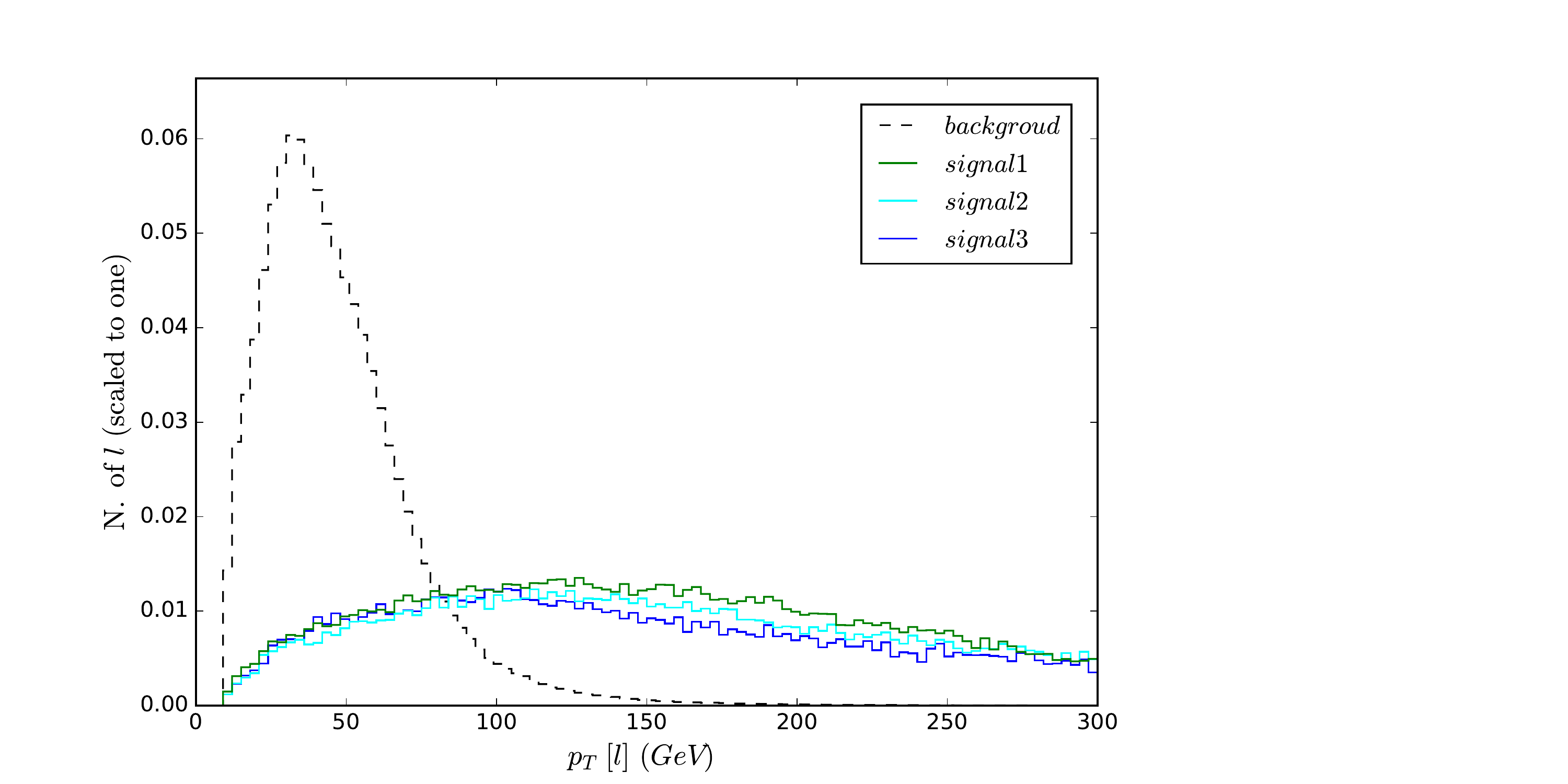}
		\end{minipage}
	}
	\centering
	\subfigure[]{
		\begin{minipage}[b]{0.48\textwidth}
			\includegraphics[width=1.45\textwidth]{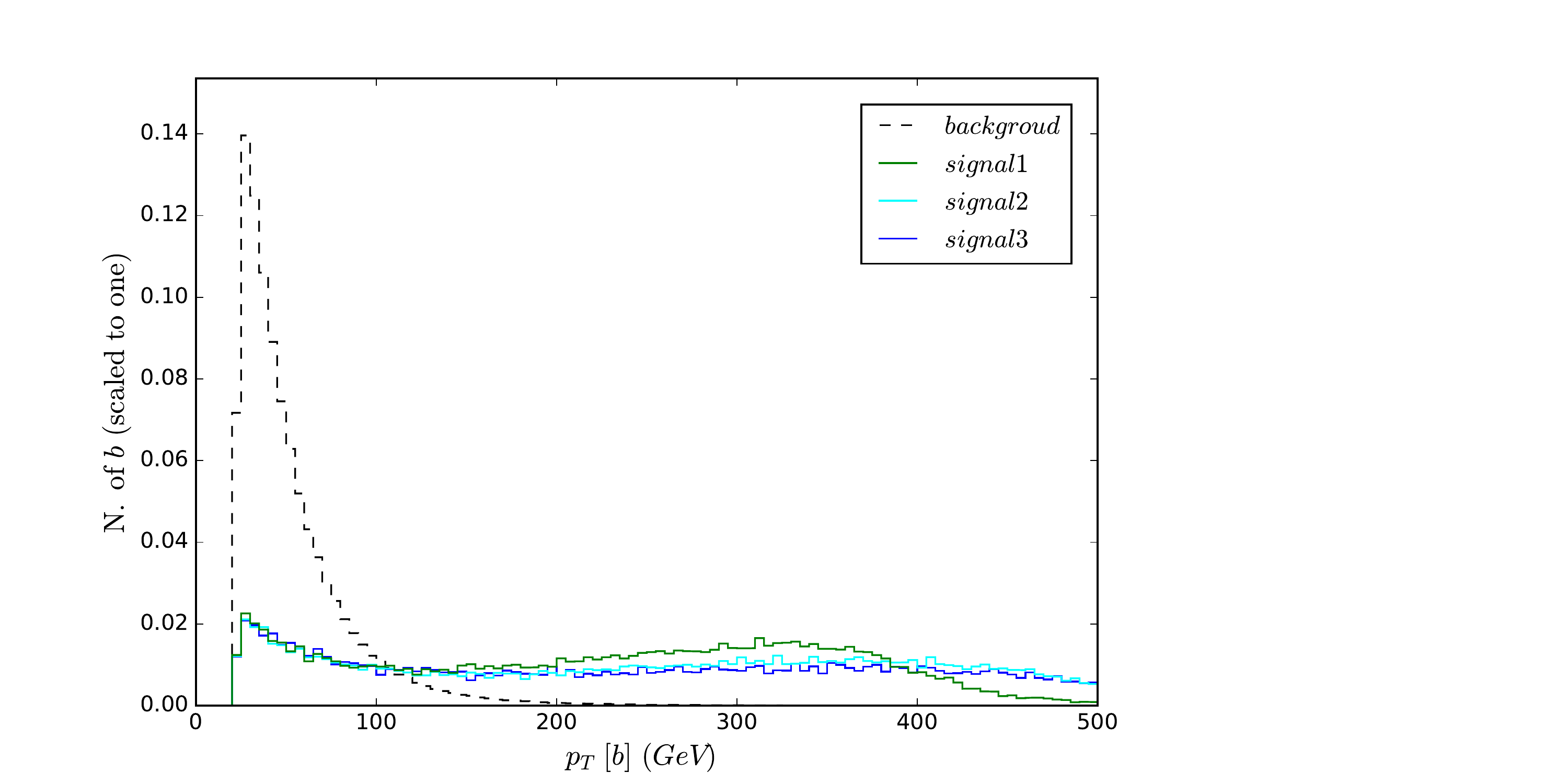}
		\end{minipage}
	}
	\centering
	\subfigure[]{	
		\begin{minipage}[b]{0.48\textwidth}
			\includegraphics[width=1.45\textwidth]{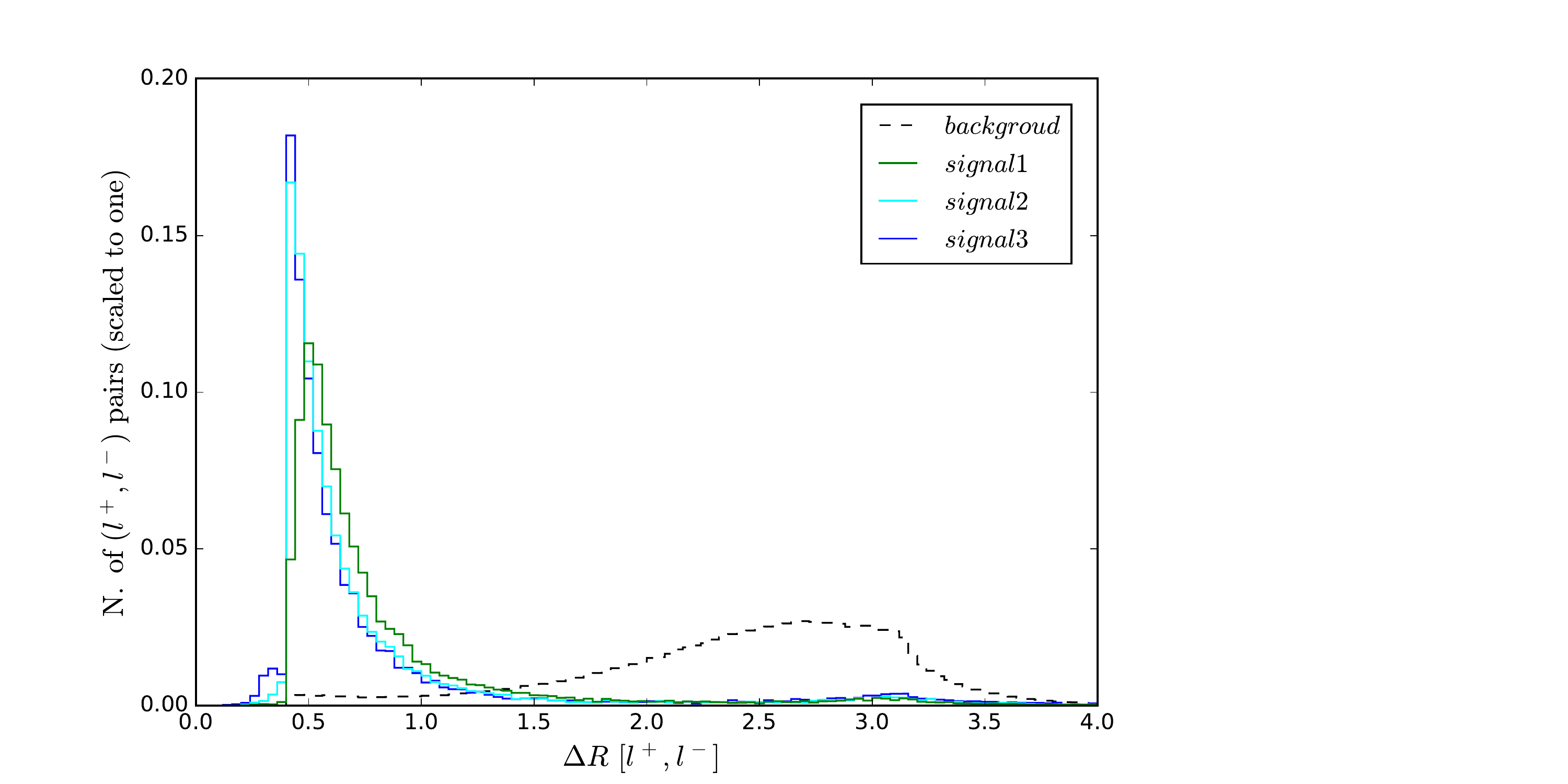}
		\end{minipage}
	}
	\centering
	\subfigure[]{	
		\begin{minipage}[b]{0.48\textwidth}
			\includegraphics[width=1.45\textwidth]{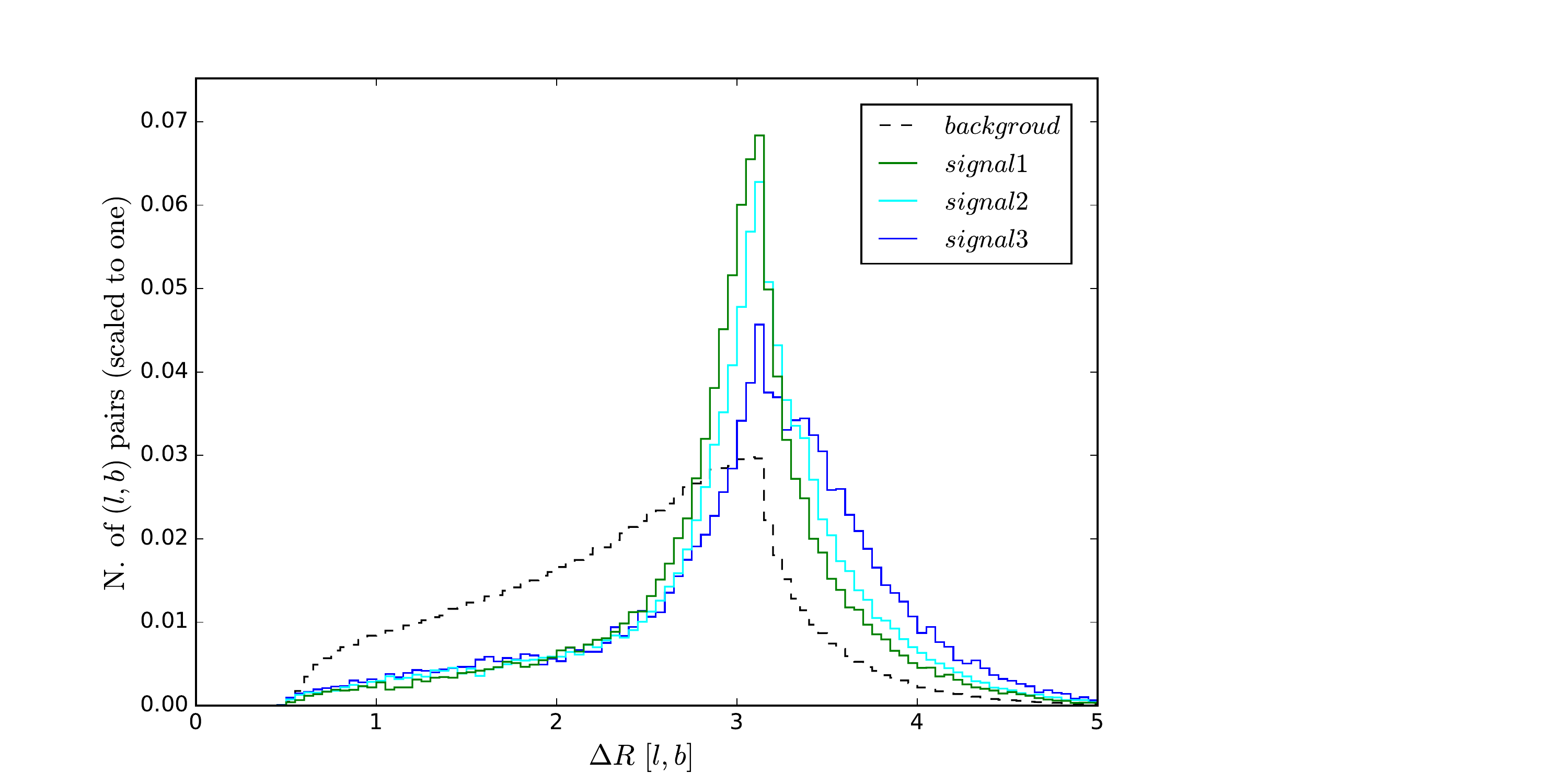}
		\end{minipage}
	}
	\centering
	\subfigure[]{	
		\begin{minipage}[b]{0.48\textwidth}
			\includegraphics[width=1.45\textwidth]{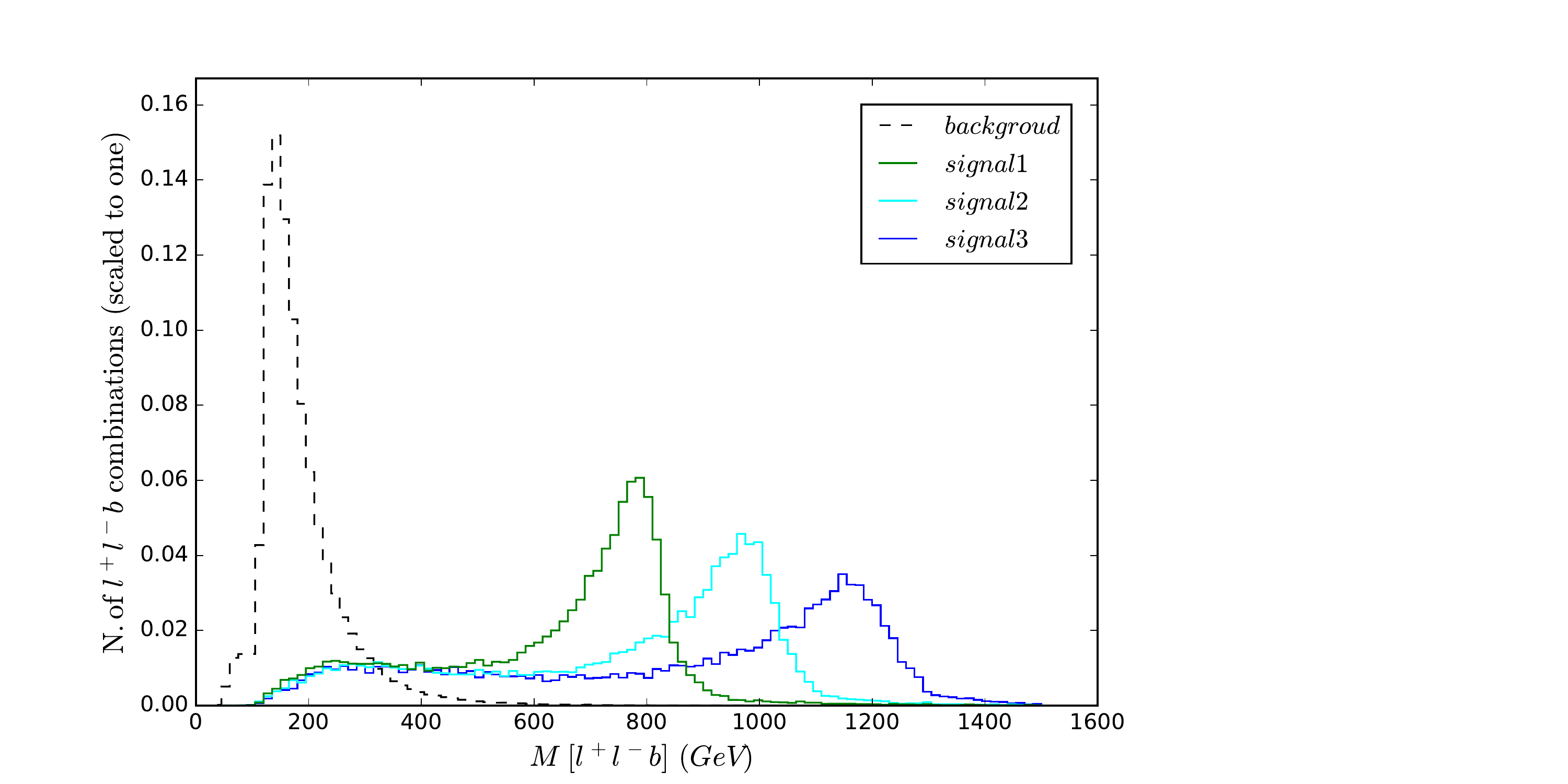}
		\end{minipage}
	}
	\caption{Normalized distributions of the transverse momentums $p^{l,b}_T$, the particle separations \hspace*{1.65cm} $\Delta R(l,b)$, $\Delta R(l^+,l^-)$, and the invariant mass $M(l^+l^-b)$ for the signals and backgrounds \hspace*{1.65cm} after basic cuts at the 14 TeV LHC with an integrated luminosity of 300~ $\rm fb^{-1}$. }
	\label{fig5}
\end{figure}

Now, we consider the signatures of VLQ-$B$ at the 14 TeV LHC through the process $pp\rightarrow B \rightarrow Zb$ with $Z$ subsequently  decaying to pair of leptons.
We analyze the observation potential by performing  Monte Carlo simulation of the signal and background events, and applying the suitable selection cuts to maximize the statistical significance. We choose the following three benchmark points: $M_{B_1}$ = 800 GeV,  $M_{B_2}$ = 1000 GeV,  $M_{B_3}$ = 1200 GeV with $\kappa^b$= 0.1, and then explore the sensitivity of VLQ-$B$ at the 14 TeV LHC through the channel
\begin{equation}
pp\rightarrow B \to Z (\rightarrow l^+l^-)b.
\end{equation}
 The main SM backgrounds are the processes containing two leptons and one $b$ jet in the final states, such as
\begin{eqnarray}
pp\to~ Z (\to l^{+}l^{-})b\;.
\end{eqnarray}

To simplify our analysis, we will assume $l^\pm= e ^\pm$ and $ \mu^\pm$. In this case, the total cross section of the main SM backgrounds is about 18.01 $\rm pb$  at the 14 TeV LHC.
The signal and background processes are simulated at the 14 TeV LHC with an integrated luminosity of 300 $\rm fb^{-1}$.
First of all, we pick up the events that include one opposite-sign same-flavor lepton pair and one $b$ jet after imposing the following basic cuts:
\begin{eqnarray}
&& p^{l}_{T}> 10~{\rm GeV} \;,\quad  p^{j}_{T}> 20~{\rm GeV} \;,\quad
\vert\eta(l)\vert<2.5 \;,\quad \vert\eta(j)\vert<5 \;.\quad
\Delta R(ll,lj)>0.4 \;.
\end{eqnarray}
Where $p^{}_T$ denotes the transverse momentum and $\eta$ is the pseudo-rapidity, the particle separation $\Delta R_{ij}$ is defined as $\sqrt{ \Delta \eta_{ij}^2 + \Delta \phi_{ij}^2 }$ with $\Delta \eta_{ij}$ and $\Delta \phi_{ij}$ being the rapidity and azimuthal angle gaps between the two particles in a pair. These basic cuts are used to simulate the geometrical acceptance and detection threshold of the detector.

In order to get some hints of further cuts for reducing the SM backgrounds, we show the normalized distributions of $p^{l,b}_T$, $\Delta R(l,b)$, $\Delta R(l^+,l^-)$ and the invariant mass $M(l^+l^-b)$ for the signals and backgrounds after the basic cuts in Fig.~\ref{fig5}, where the signals~1, 2 and 3 correspond to $M_{B_1}$ = 800 GeV,  $M_{B_2}$ = 1000 GeV and $M_{B_3}$ = 1200 GeV, respectively. As shown in Fig.~\ref{fig5}, because of the larger mass of VLQ-$B$, the decay products of VLQ-$B$ are highly boosted. Therefore, the $p^{l,b}_T$ peaks of the signals are larger than those of the SM backgrounds and the lepton pairs of the signal are much closer.
For the distribution of the invariant mass $M(l^+l^-b)$, the signal events have peaks around the VLQ-$B$ mass, while the background events do not. According to the information of these kinematic distributions, we impose the following cuts to get a high statistical significance:
\begin{eqnarray}
\rm{Cut\!-\!1}&:& p^{b}_{T} > 150~{\rm GeV};\nonumber
\\\rm{Cut\!-\!2}&:& p^{l}_{T} > 100~{\rm GeV};\nonumber
\\\rm{Cut\!-\!3}&:& \Delta R(l^+l^-) < 1.0;\nonumber
\\\rm{Cut\!-\!4}&:& \Delta R(l,b) > 3.0;\nonumber
\\\rm{Cut\!-\!5}&:& M(l^+l^-b) > 700 ~{\rm GeV}.\nonumber
\end{eqnarray}

In order to see whether the signatures of VLQ-$B$ can be detected at the 14 TeV LHC, we further calculate the statistical significance of signal events $SS$= $S/\sqrt{S+B}$, where $S$ and $B$ denote the numbers of the signal and background events, respectively. Our numerical results are shown in Table~\ref{table1}.
\begin{table*}[htbp]
	\caption{ Effects of individual kinematic cuts on the signals and backgrounds for $M_{B_1}$ = 800GeV, \hspace*{1.65cm}$M_{B_2}$ = 1000GeV,  $M_{B_3}$ = 1200 GeV with $\kappa^b$=0.1.}
	\hspace{15cm}
	\centering
	\begin{tabular}{|c|c|c|c|c|c|c|c|}
		\hline
		\multicolumn{8}{|c|}{ LHC @14~TeV, ~$\mathcal{L}$ = 300 $\rm fb^{-1}$}\\
		\hline
		\multicolumn{1}{|c|}{}
		&\multicolumn{3}{c|}{Signals(S)}
		&\multirow{2}*{Background(B)}
		&\multicolumn{3}{c|}{\multirow{1}*{$S/\sqrt{S+B}$}}\\
		\cline{2-4}
		\cline{6-8}
		&$M_{B_1}$&$M_{B_2}$&$M_{B_3}$&&$M_{B_1}$&$M_{B_2}$&$M_{B_3}$\\
		\hline
		Basic cuts  & 607.4 &147.4&36.3 &$1.31\times10^6$&0.53&0.13&0.03 \\
		\hline
		$\rm{Cut\!-\!1}$  & 493.1&121.3&29.2&$2.47\times10^4$&3.11&0.77&0.19\\
		\hline
		$\rm{Cut\!-\!2}$& 486.6  &120.2&29.0&$1.92\times10^4$&3.46&0.86&0.21 \\
		\hline
		$\rm{Cut\!-\!3} $ & 449.9&114.1&27.6&6385.2&5.44&1.42&0.34 \\
		\hline
		$\rm{Cut\!-\!4} $&378.6&101.5&24.9 &4312.1&5.53&1.53&0.38 \\
		\hline
		$\rm{Cut\!-\!5} $&294.0 &96.9&24.4&576.9&9.96&3.73&0.996 \\
		\hline
	\end{tabular}
	\label{table1}
\end{table*}
\begin{figure}[htbp]
	\centering
	\subfigure{
		\begin{minipage}[b]{0.47\textwidth}
			\includegraphics[width=1\textwidth]{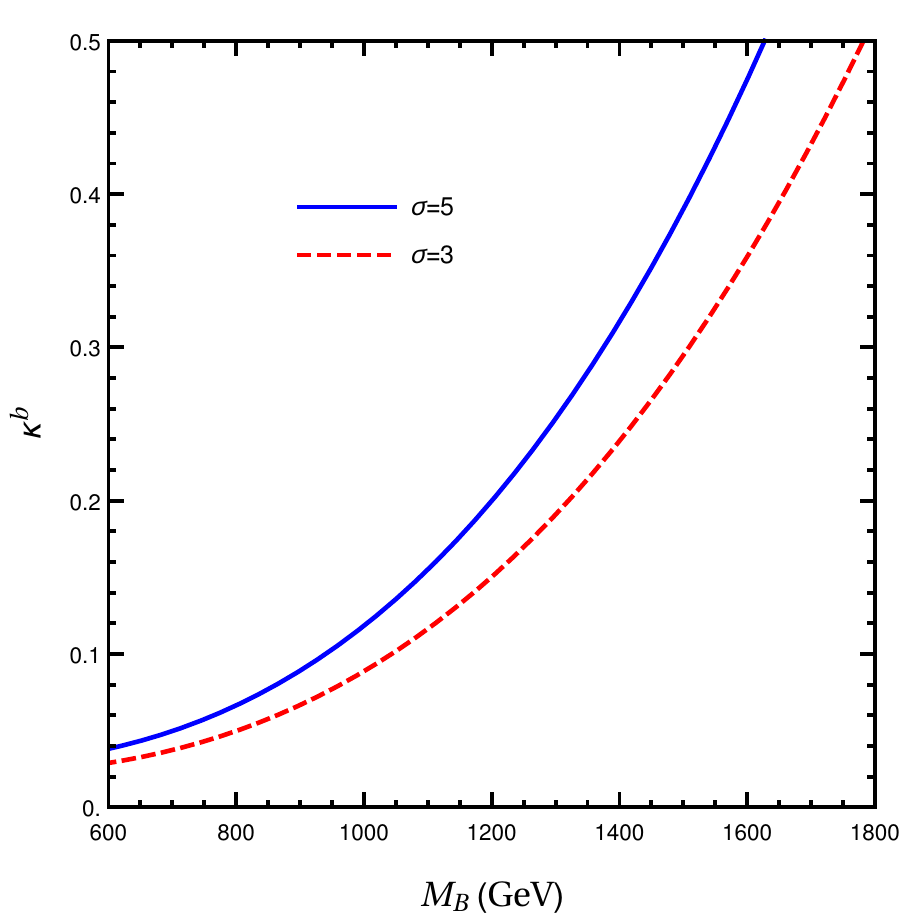}
		\end{minipage}
	}
	\caption{ 3$\sigma$ and 5$\sigma$  curves in the $M_B$-$\kappa^b$ plane for the process $pp\rightarrow B \rightarrow Zb$ at the 14 \hspace*{1.65cm} TeV LHC with the integrated luminosity $\mathcal{L}$ = 300 $\rm fb^{-1}$.}
	\label{fig6}
\end{figure}

Based on the above benchmark points, the exclusion limits of the free parameters $M_{B}$ and $\kappa^{b}$ are shown in Fig.~\ref{fig6}.
We can see that the values of $\kappa^b$  can respectively reach about 0.05~(0.07), 0.10~(0.12), and~0.15~(0.20) for $M_B$=800, 1000, and 1200 GeV at 3$\sigma$~(5$\sigma$) level at the 14 TeV LHC with the integrated luminosity $\mathcal{L}$ = 300 $\rm fb^{-1}$. The detectable lower limits of the parameter $\kappa^b$ enhance with $M_B$ increasing. So, we can say that, with an upgraded energy and higher luminosity, the signals of VLQ-$B$ might be detected through the process $pp\rightarrow B \rightarrow Zb$ at the LHC in near future.

\section*{IV. Conclusions}

VLQs have been presented in many new physics models beyond the SM, which can be produced singly and in pairs at the LHC. When the mass of VLQ is large enough, the single production processes become important, which can not only be used to detect the possible signals of VLQ, but also be used to test the mixing between the VLQ and SM quark. In this letter, we first discuss all possible decay modes and  single production channels of the $SU(2)$ singlet VLQ-$B$ at the LHC in a model independent way. Then we investigate its observability via the process $pp \to B \to Z~(\to l^+ l^-) b$ at the 14 TeV LHC. Our numerical results show that the production cross section of the process $pp \to B \to Z b$ strongly depends on the free parameters  $\kappa^b$ and  $M_B$. Its value is in the range of $3.09\times10^{-4} \sim 276.72~ \rm pb$ for $0 \leq \kappa^b \leq 0.5$ and 300 GeV $<M_B<$ 2000 GeV. We focus our attention on analysis of the signals of VLQ-$B$ through the $l^+l^-b$ final state, where $l^\pm$ are assumed to be $ e ^\pm$ and $\mu^\pm$. After simulating the signals as well as the SM backgrounds and applying suitable kinematic cuts on the relevant variables, the signal significance $SS$ can reach 9.96 and 0.996 for $\kappa^b=0.1$, $M_B$=800 and 1000GeV, respectively.  Certainly, the value of $SS$  enhances with  $\kappa^b$ increasing for the fixed value of $M_B$. We further study the detectable lower limits of the free parameters $M_{B}$ and $\kappa^{b}$ and obtain the minimum values of the free parameters $\kappa^b$ and $M_B$ with the integrated luminosity required for the discovery of such kind of down-type VLQ. We find that the values of $\kappa^b$ can respectively reach about 0.05~(0.07), 0.10~(0.12), and~0.15~(0.20)  for $M_B$=800, 1000, and 1200 GeV at 3$\sigma$~(5$\sigma$) level  at the 14 TeV LHC with the integrated luminosity $\mathcal{L}$ = 300 $\rm fb^{-1}$. Thus, we expect that the possible  signals of the VLQ-$B$ with electric charge of $Q_B= -1/3$, which is  the $SU(2)$ singlet, could be detected via the process $pp \to B \to Z~(\to l^+ l^-) b$  at the upgraded LHC.

\section*{ACKNOWLEDGEMENT}
This work was partially supported by the National Natural Science Foundation of China under Grants No. 11875157, Grant No. 11847303 and Grants No.11847019.

\end{document}